\documentclass[twocolumn]{aastex631}

\usepackage{amsfonts}
\usepackage{amssymb}
\usepackage{amsmath}

\usepackage{xcolor}

\usepackage{comment}

\newcommand{\result}[1]{#1}

\begin{document}

\title{
Selection Effects in Periodic X-ray Data from Maximizing Detection Statistics
}

\author{Reed Essick}
\email{ressick@perimeterinstitute.ca}
\affiliation{Perimeter Institute for Theoretical Physics, 31 Caroline Street North, Waterloo, Ontario, Canada, N2L 2Y5}

\begin{abstract}
    The Neutron Star Interior Composition Explorer (NICER) records data of exceptional quality on the energy-dependent X-ray pulse profile of pulsars.
    However, in searching for evidence of pulsations,~\citet{Guillot:2019} introduce a procedure to select an ordered subset of the data that maximizes a detection statistic (the H-test).
    I show that this procedure can degrade subsequent analyses using an idealized model with a stationary expected count rates from both noise and signal.
    Specifically, the data-selection procedure biases the inferred mean count rate to be too low, biases the inferred pulsation amplitude to be too high, and that the size of these biases scales strongly with the amount of data that is rejected and the true signal amplitude.
    The procedure also alters the null-distribution of the H-test rendering nominal detection significance estimates overly optimistic.
    While the idealized model does not capture all the complexities of real NICER data, it suggests that these biases could be important for NICER's observations of J0740+6620 and other faint pulsars (observations of J0030+0451 are likely less affected).
    I estimate that these selection effects may introduce a bias of \result{$\mathcal{O}(10\%)$} on average in the inferred modulation depth of lightcurves like J0740+6620's, and may be as large as \result{$\mathcal{O}(50\%)$} for fainter pulsars.
    However, the change for a single dataset like J0740+6620 is expected to be a shift between \result{-5\% and +20\%}.
    This could imply that the lower limit on J0740+6620's radius is slightly larger than it should be, although preliminary investigations suggest the change in the radius constraints are \result{$\mathcal{O}(1\%)$} with real J0740+6620 data.
\end{abstract}

\section{Introduction}
\label{sec:introduction}

Over the past few years, the Neutron Star Interior Composition Explorer~\citep[NICER;][]{nicer:designanddevelopment, nicer:descriptionandperformance} released simultaneous measurements of the mass and radius of two X-ray pulsars: J0030+0451~\citep{miller:2019, riley:2019} and J0740+6220~\citep{wolff:2021, miller:2021, riley:2021, raaijmakers:2021}.
These observations inform our understanding of Neutron Star (NS) composition and structure, primarily through inferred constraints on the Equation of State (EoS) of extremely dense matter.
Together with Gravitational Wave (GW) observations of the coalescence of compact binaries containing NSs, such as the observation of GW170817~\citep{gw170817detection, gw170817equationofstate, gw170817properties} with the advanced LIGO~\citep{ligo} and Virgo~\citep{virgo} interferometers, NICER X-ray observations set some of the most stringent constraints available to-date on the properties of matter at supranuclear densities ($\gtrsim 2.8\times10^{14}\,\mathrm{g}/\mathrm{cm}^{3}$).
See, e.g.,~\citet{chatziioannou:2020} for a review.

Several authors have also remarked on the fact that the inferred radii from NICER observations tend to lie at the upper end of what is consistent with constraints from GW170817.
For example, Fig. 3 in~\citet{landry:2020} and Fig. 3 in~\citet{legred:2021} both show the inferred masses and radii of individual objects from X-ray observations when considered separately and when considered jointly with GW data.
For both pulsars, the X-ray data alone tend to prefer larger radii, as reported.
This may just be a statistical fluctuation, as the posteriors are still broad, there are only a handful of events, and there is no real tension between GW and X-ray observations.
However, as I will show, NICER's current data-selection procedure may introduce an artificial preference for larger radii.
To the best of my knowledge, this effect has not been previously described in the literature.
As the number of observed systems continues to grow, it will be important to quantify all possible selection effects and associated biases if we are to obtain precise (and accurate!) constraints on the EoS.

In particular, recent constraints on J0740+6620 (the most massive NS with a confident mass measurement to-date) suggest $R\gtrsim13\,\mathrm{km}$ for $M\simeq2.08\,M_\odot$~\citep{miller:2021}.
This is a relatively large value, and it is of interest to know how robust the lower limit for the radius is.
\citet{miller:2021} discuss this in some detail, concluding that their lower limit should not change much due to possible systematic errors.
However, they do not consider the possible impact of selection effects within the data-selection procedure.

This Letter's aim is to explain why there may be need for concern with current NICER procedures by examining an idealized model.
I stress that this does not prove there are issues with any of the published constraints, but hope that it motivates further study to ensure the NICER observations can be utilized to their full potential.

I begin in Sec.~\ref{sec:review of the h-test} by reviewing the detection statistic employed by NICER to identify periodic signals: the H-test~\citep{dejager:1989, dejager:2010}.
I then review the data-selection procedure adopted by the NICER collaboration for J0740+6620 as described in~\citet{Guillot:2019} and~\citet{wolff:2021}.
In particular, I examine the possible impact of the choice to retain only the subset of data that maximizes the H-test.
For example, \citet{dejager:1989} note that the H-test is more sensitive to signals with larger relative oscillation amplitudes and narrower beaming patterns.
By selecting the data that maximizes the significance of the inferred periodicity in the (unknown) light-curve, the inferred size of oscillations in the same signal may be artificially inflated.
In turn, this could lead to a smaller inferred compactness and larger inferred radius at a fixed mass.
Sec.~\ref{sec:toy model} investigates this quantitatively within the context of an idealized model.
I find that~\citet{Guillot:2019}'s data-selection procedure biases the inferred signal parameters to smaller phase-averaged rates and larger modulation depths.
These effects are larger for weaker signals.
The data-selection procedure also alters the null-distribution of the H-test, implying that standard significance estimates cannot be used.
I summarize my findings in Sec.~\ref{sec:discussion}, quantitatively estimating the implications for current measurements of J0030+0451 and J0740+6620 in turn.

\section{The H-test for Weak, Periodic Signals}
\label{sec:review of the h-test}

I first review the H-test and its use in detecting periodic signals with unknown light-curves in sparse X-ray observations (fundamental period of the light-curve is much shorter than the average time between events).
Readers should refer to~\citet{dejager:1989, dejager:2010} for more details.

The H-test provides a robust null-test for uniformity on a circle and, as such, is particularly useful when searching for the presence of periodic features in phase-folded X-ray data.
\citet{dejager:1989} showed that the H-test is typically as sensitive or more sensitive than other tests when the signal is small and the light curve is unknown.
The H-test works by considering deviations in the light-curve ($f(\phi)$) from a uniform distribution, defining
\begin{equation}
    \psi(f) \equiv \int\limits_0^{2\pi}d\phi\, \left(f(\phi)-\frac{1}{2\pi}\right)^2
\end{equation}
and modeling the observed light-curve with a Fourier Series Estimator with $m$ harmonics
\begin{equation}
    \hat{f}_m = \frac{1}{2}\left( 1 + \sum\limits_{k=1}^m \left[ \hat{\alpha}_k \cos(k\phi) + \hat{\beta}_k \sin(k\phi) \right] \right)
\end{equation}
with the Fourier coefficients estimated via Monte Carlo sums over the set of $n$ observed phase data $\{\phi_i\}$
\begin{align}
    \hat{\alpha}_k & = \frac{1}{n}\sum_i^n \cos(k\phi_i) \\
    \hat{\beta}_k & = \frac{1}{n}\sum_i^n \sin(k\phi_i)
\end{align}
For a given harmonic $m$, we define the deviation statistic for a set of phase observations
\begin{equation}
    Z^2_m(\{\phi_i\}) = 2\pi n \psi(\hat{f}_m) = 2n\sum\limits_{k=1}^m \left[ \hat{\alpha}_k^2 + \hat{\beta}_k^2 \right]
\end{equation}
The H-test is then defined as a maximization over different harmonics
\begin{equation}
    H(\{\phi_i\}) = \max\limits_{1 \leq m \leq 20} \left[ Z^2_m(\{\phi_i\}) - 4m + 4 \right]
\end{equation}
If $H$ is sufficiently large, then the null hypothesis that the data are uniformly distributed (no periodicity) is rejected, and the presence of a periodic signal is inferred.
\citet{dejager:2010} find that the survival function for $H$ is exponential:
\begin{equation}\label{eq:nominal pvalue}
    P(H>H_\mathrm{obs}) = \exp\left(-0.4H_\mathrm{obs}\right) .
\end{equation}

Additionally, \citet{dejager:1989} discuss possible selection effects associated with the H-test.
They note that the H-test will tend to favor sources that have deeper relatively oscillation amplitudes and narrower radiation beams (lower pulse duty cycles).
In the context of NICER's observations, both effects will tend to prefer less compact stars (less self-lensing implies deeper oscillations and narrower beams).
Selecting subsets of data based on what maximizes the H-test may introduce a preference for data that minimizes the inferred stellar compactness (maximizes the inferred radius at a fixed mass).

\section{Review of NICER's Data-Selection Procedure}
\label{sec:review of data selection}

\begin{figure*}
    \includegraphics[width=1.0\textwidth]{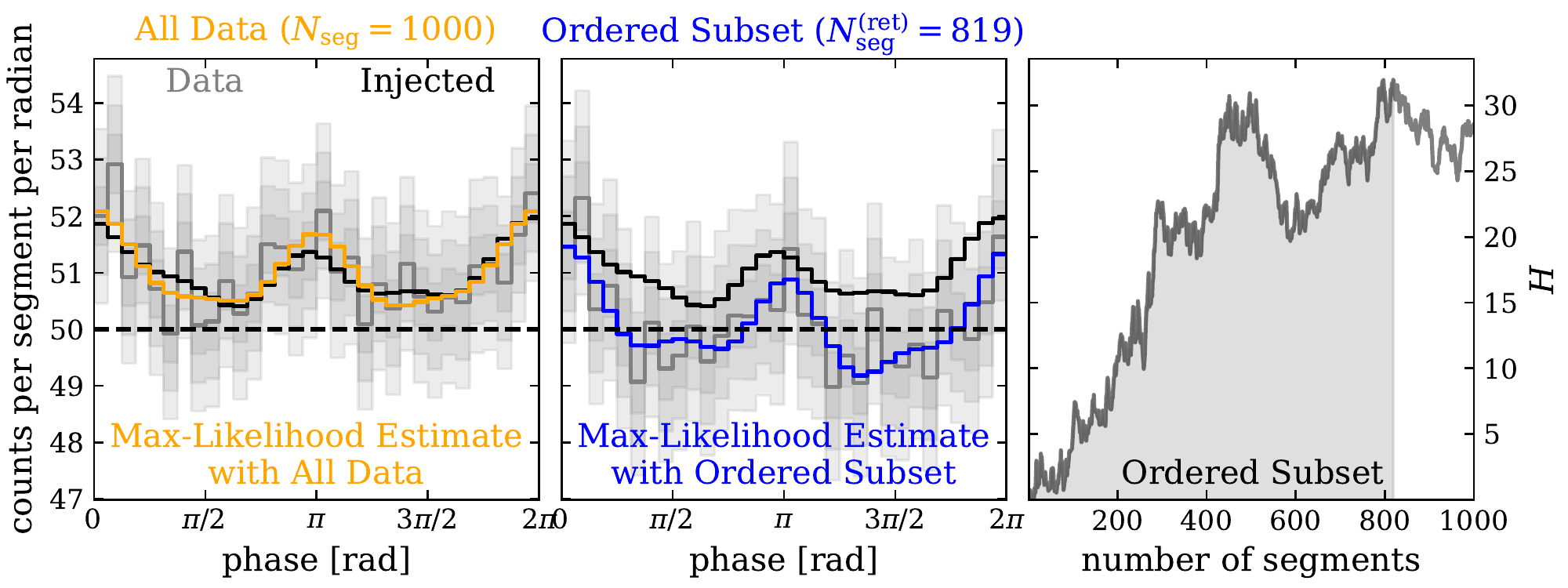}
    \caption{
        Examples of binned (\emph{grey}) data, (\emph{orange/blue}) maximum-likelihood estimates for the signal, and (\emph{black}) true injection when we consider (\emph{left}) all available data ($N_\mathrm{seg}=1000$) and (\emph{middle}) the selected subset of data following~\citet{Guillot:2019}'s procedure ($N_\mathrm{seg}^{(\mathrm{ret})} = 819$).
        (\emph{solid black}) Total and (\emph{dashed black}) noise-only differential rates from the true signal.
        Shaded regions approximate 1-, 2-, and 3-$\sigma$ Poisson uncertainties on the true rate based on the observed data.
        I use 32 phase bins and $\tau a_0 = 1$.
        (\emph{right}) The corresponding trajectory of the cumulative H-test statistic ($H$) as additional segments are included in order of increasing counts.
        The subset of data that maximizes $H$ is shaded.
    }
    \label{fig:binned models and data}
\end{figure*}

NICER is sensitive to soft X-rays (0.2--12 keV) and provides both exceptional energy and timing resolution~\citep{nicer:designanddevelopment, nicer:descriptionandperformance}.
Typical timing uncertainty is $<100\,\mathrm{ns}$, which is much less than the rotation period of any observed pulsar.
This provides a measurement of the energy-dependent pulsation pattern of X-ray sources by directly measuring the energy and time-of-arrival of individual photons.
The instrument is mounted on the International Space Station (ISS) and, given its environment (location of the ISS, space weather, etc.) and current target (pointing with respect to the Sun, Earth, and Moon, etc.), is subject to a variety of possible noise sources.
The noise sources can vary over time and, as such, not all data recorded by NICER are equivalent.
For this reason, the NICER team divides the total observation into small segments~\citep[each spanning 10--100 sec,][]{Guillot:2019}, calling each segment a \emph{good time interval} (GTI).
As described in~\citet{Guillot:2019}, the final set of GTI's included in downstream analyses is chosen by maximizing the H-test significance.

\citet{Guillot:2019} define a procedure for ordering GTI's by rough estimates of their backgrounds (GTI's with smaller estimated backgrounds first) and include only those GTI's that increase the cumulative H-test significance (see, e.g., their Fig. 1).
Specifically, they assume that the total count rate is dominated by the background (signals are weak) and order GTI's by the ratio of their total number of counts and their durations.
GTI's are then included cumulatively until the corresponding value of $H$ is maximized.
\citet{Guillot:2019} advocate discarding all remaining data and only retaining the ordered subset that maximizes $H$.
Fig.~\ref{fig:binned models and data} shows an example trajectory of $H$ in our idealized model as more data are included, with a maximum reached with only \result{$\approx 80\%$} of the data.

This ordering is intended to include GTI's with smaller intrinsic background rates first.
However, I show in Sec.~\ref{sec:toy model} that this procedure causes several issues even in idealized situations.
It is reasonable to expect these issues persist in more complicated situations, although further study is needed.

\citet{Guillot:2019} additionally perform a grid search over which energies are considered (between 0.2--2.0 keV) in order to further maximize the detection significance (see their Table 3).
I do not consider the implications of such an optimization in this work, as I \textit{de facto} consider a single energy channel.
However, it is likely that optimization over the energy channels could introduce additional biases.

\section{Data Selection for Sparse X-ray Observations}
\label{sec:toy model}

I now consider an idealized Poisson model for sparse X-ray observations.
That is, I consider differential Poisson rates for the signal and background as a function of rotation phase
\begin{align}
    \lambda_n & = \mathrm{constant} \\
    \lambda_s & = f(\phi) \geq 0 \nonumber \\
              & = a_0 + \sum_{m=1}^M a_m \cos\left( m\phi + \delta_m \right)
\end{align}
where I assume the noise is uncorrelated with the rotation phase and express the signal as a periodic function with $M$ harmonics.
Furthermore, the expected rate of counts of type $i$ in the interval [$\phi_1$, $\phi_2$) is
\begin{equation}
    \Lambda_i(\phi_1, \phi_2) = \int\limits_{\phi_1}^{\phi_2} d\phi\, \lambda_i(\phi)
\end{equation}
and the corresponding expected number of counts within the same phase interval throughout an observation with duration $T$ is
$\Lambda_i(\phi_1, \phi_2) T$.
We then model the likelihood for $N_\mathrm{phs}$ binned phase measurements via
\begin{align}
    \log p(\{c_i\}|\lambda_n, \lambda_s) & = \sum\limits_i^{N_\mathrm{phs}} \left[ c_i \log\left(\Lambda(\phi_i, \phi_{i+1})T\right) - \log\left(c_i!\right) \right] \nonumber \\
        & \ \ \quad - \Lambda(0, 2\pi) T \label{eq:likelihood}
\end{align}
where $c_i$ is the number of observed counts within $[\phi_i, \phi_{i+1})$, $\phi_0=0$, $\phi_{N_\mathrm{phs}+1}=2\pi$, and 
\begin{equation}
    \Lambda(\phi_1, \phi_2) = \Lambda_n(\phi_1, \phi_2) + \Lambda_s(\phi_1, \phi_2)
\end{equation}
is the total expected rate of events between $\phi_1$ and $\phi_2$ from both signal and noise.

Furthermore, I consider a sequence of $N_\mathrm{seg}$ independent, identically distributed (i.i.d.) segments, each with duration $\tau$ such that $T = \tau N_\mathrm{seg}$.
I proceed by generating random realizations of observed data from this model separately for each segment and then consider the impact on the global inference when I employ various data-selection algorithms.

\begin{figure*}
    \includegraphics[width=1.0\textwidth]{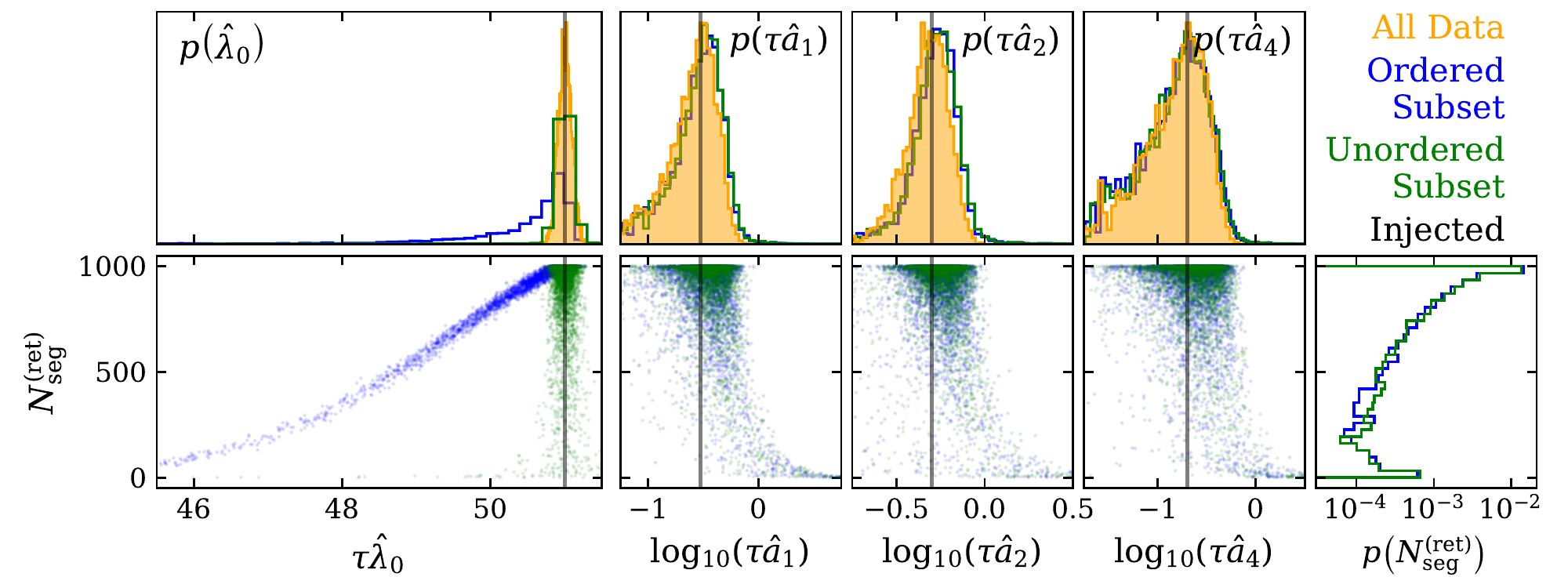}
    \caption{
        (\emph{top}) Marginal distributions for estimators of signal parameters and (\emph{bottom}) joint distributions of estimators with the number of segments retained by H-test maximization.
        I show distributions of \result{4000} realizations containing \result{1000} segments each with \result{$\tau a_0 = 1$} when using (\emph{orange, shaded}) all data, (\emph{blue, unshaded}) the retained subset of segments ordered by counts, (\emph{green, unshaded}) the retained subset of randomly ordered segments, and (\emph{black}) the injected parameters.
        There is a strong correlation between $N_\mathrm{seg}^{(\mathrm{ret})}$ and $\hat{\lambda}_0$ and weaker correlations with $\hat{a}_{m}$.
    }
    \label{fig:correlations}
\end{figure*}

Throughout the following, I assume $\tau = 100\,\mathrm{sec}$ for all segments and \result{$\tau\lambda_n = 50$}.
I investigate various signal strengths ranging between \result{$\tau a_0 = 0.1$} and \result{$10$}.
On average, then, each segment contains \result{$\mathcal{O}(2\pi\cdot50)$} counts.
For simplicity, I consider a single injected light-curve with three non-zero harmonics
\begin{align}
    a_1 / a_0 = 0.3 & , \ \delta_1 = 0.0 \\
    a_2 / a_0 = 0.5 & , \ \delta_2 = 0.0 \\
    a_4 / a_0 = 0.2 & , \ \delta_4 = 0.75
\end{align}
and scale $a_0$ to control the overall signal strength.
Fig.~\ref{fig:binned models and data} shows an example realization with \result{$N_\mathrm{seg} = 1000$} and \result{$\tau a_0 = 1$}.

\subsection{Biases and Correlations}
\label{sec:biases and correlations}

To begin, I consider the impact of the data selection process on the inferred signal parameters by computing maximum-likelihood estimates (MLEs) of the signal parameters for many realizations of our experiment.
For computationally expediency, I assume $a_m \ll \lambda_n + a_0 \ \forall \ m \geq 1$ and expand Eq.~\ref{eq:likelihood} to second order in $a_{m}$, which allows me to analytically determine MLEs for $\lambda_0 = \lambda_n + a_0$, $a_{m}\cos(\delta_m)/\lambda_0$, and $a_m\sin(\delta_{m})/\lambda_0$.
I report $\hat{a}_m \geq 0$ derived from the MLEs in this alternate parametrization.

There is a strong correlation between the number of segments retained after the data-selection procedure ($N_\mathrm{seg}^{(\mathrm{ret})}$) and $\hat{\lambda}_0$.
That is, I infer smaller $\hat{\lambda}_0$ if I retain fewer segments.
This makes sense, as~\citet{Guillot:2019}'s data selection procedure first orders segments by increasing average count rate.
Therefore, unless all the data are included, the procedure preferentially selects the data with fewer counts.
This is strongly correlated with $\hat{\lambda}_0$ as
\begin{equation}
    \hat{\lambda}_0 = \frac{1}{2\pi}\sum\limits_i c_i
\end{equation}
is directly proportional to the total number of counts.
Fig.~\ref{fig:binned models and data} demonstrates this with binned phase data from a single experiment.
Fig.~\ref{fig:correlations} shows the correlations between $N_\mathrm{seg}^{(\mathrm{ret})}$ and the inferred lightcurve model for many experiments.

Fig.~\ref{fig:correlations} demonstrates additional correlations between $N_\mathrm{seg}^{(\mathrm{ret})}$ and $\hat{a}_{m}$.
As fewer segments are retained, one tends to systematically overestimate $a_{m}$.
This can be interpreted as the H-test's preference for subsets of data that maximize the signal's modulation.
That is, when $N_\mathrm{seg}^{(\mathrm{ret})}$ is small, both the numerator and the denominator in the ratio $\hat{a}_{m}/\hat{\lambda}_0$ are biased, making the ratio larger.
This ratio defines the relative size of the oscillations within the light-curve.
Furthermore, the H-test's estimate for the optimal number of harmonics to include is often larger than the true number of harmonics when $N_\mathrm{seg}^{(\mathrm{ret})}$ is low.
Maximizing the H-test, then, selects data that appear to contain signals that are both larger and more complex than the true signal.

In line with this picture, I do not observe a large bias in $\hat{\delta}_{m}$.
Even though particular combinations of phases may make the light-curve narrower and therefore easier to detect, this does not appear to be easily achieved with random noise realizations.
As such, $\delta_{m}$ does not affect the detectability of the signal as much as $a_{m}$.

Because I find a strong correlation between the inferred mean count rate and the amount of data retained when segments are first ordered by the number of counts they contain, I also explore a possible alternative in Fig.~\ref{fig:correlations}.
I consider first randomly ordering segments (as opposed to by increasing count) and then cumulatively including segments until $H$ is maximized.
While the bias in $\hat{\lambda}_0$ is lessened (except in a few extreme cases), I still observe a bias in $\hat{a}_{m}$ that scales with $N_\mathrm{seg}^{(\mathrm{ret})}$ (compare ordered and unordered subsets in Fig.~\ref{fig:correlations}).

\begin{figure*}
    \begin{minipage}{0.5\textwidth}
        \includegraphics[width=1.0\textwidth, clip=True, trim=0cm 1.10cm 0cm 0cm]{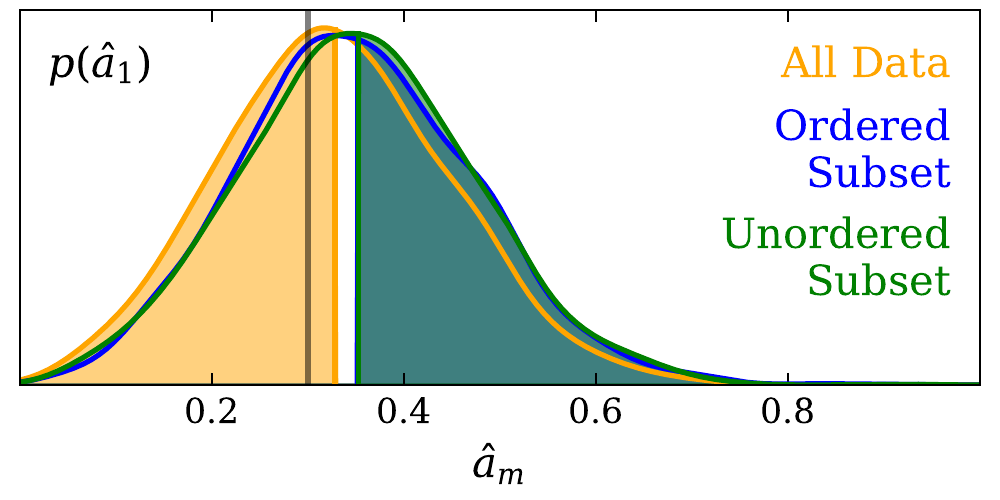}
        \includegraphics[width=1.0\textwidth, clip=True, trim=0cm 1.10cm 0cm 0cm]{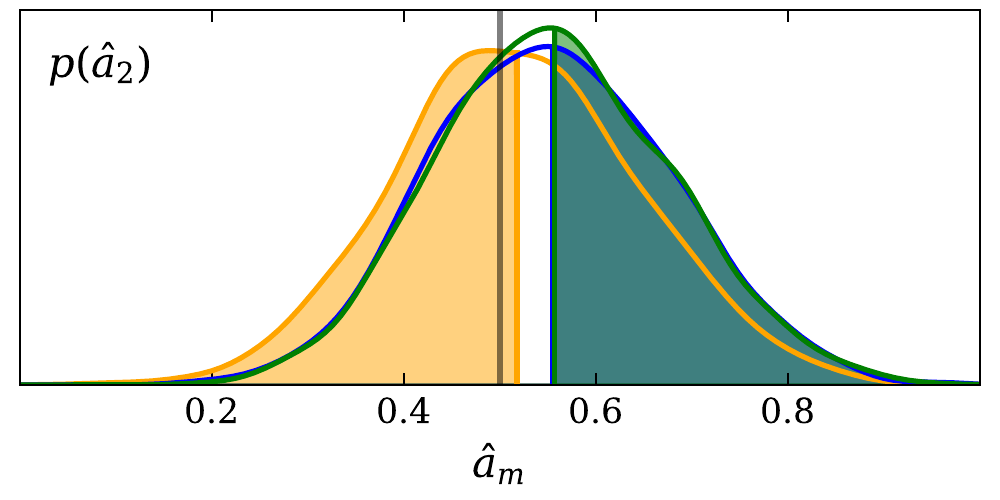}
        \includegraphics[width=1.0\textwidth]{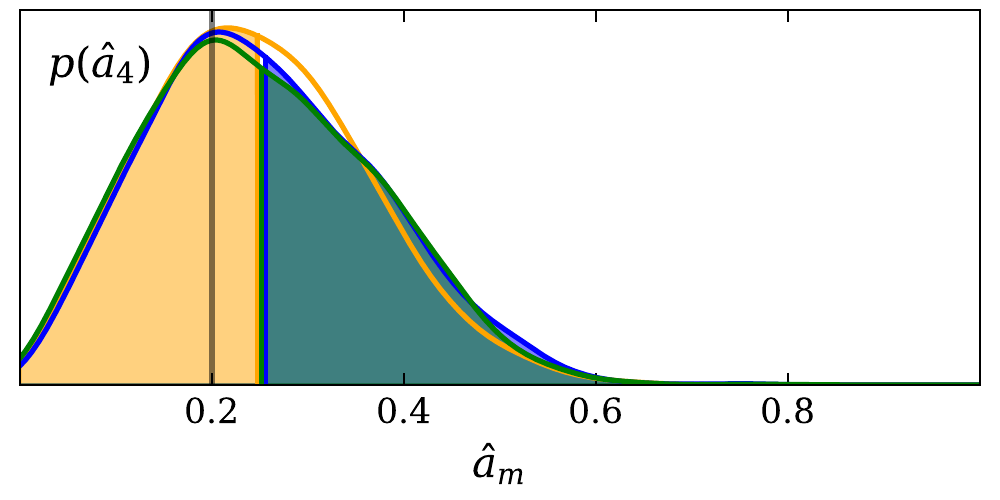}
    \end{minipage}
    \begin{minipage}{0.5\textwidth}
        \includegraphics[width=1.0\textwidth, clip=True, trim=0cm 1.10cm 0cm 0cm]{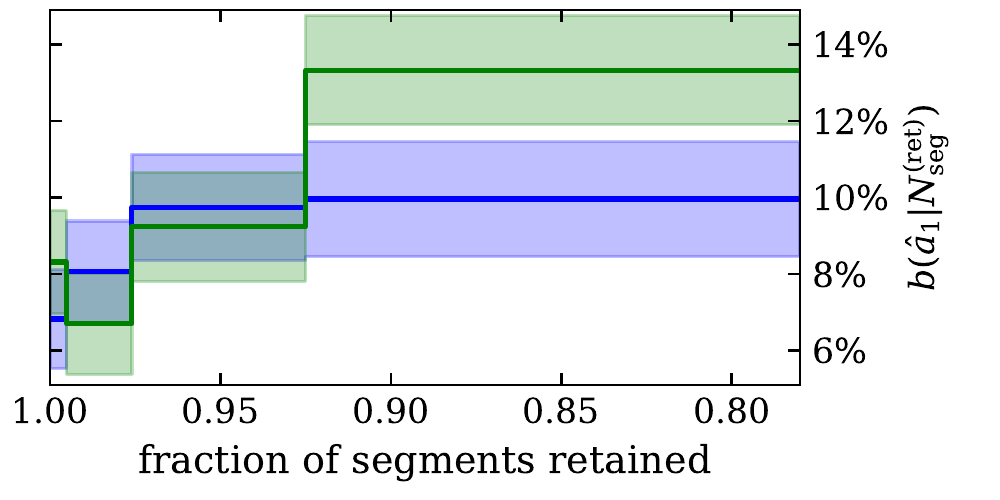}
        \includegraphics[width=1.0\textwidth, clip=True, trim=0cm 1.10cm 0cm 0cm]{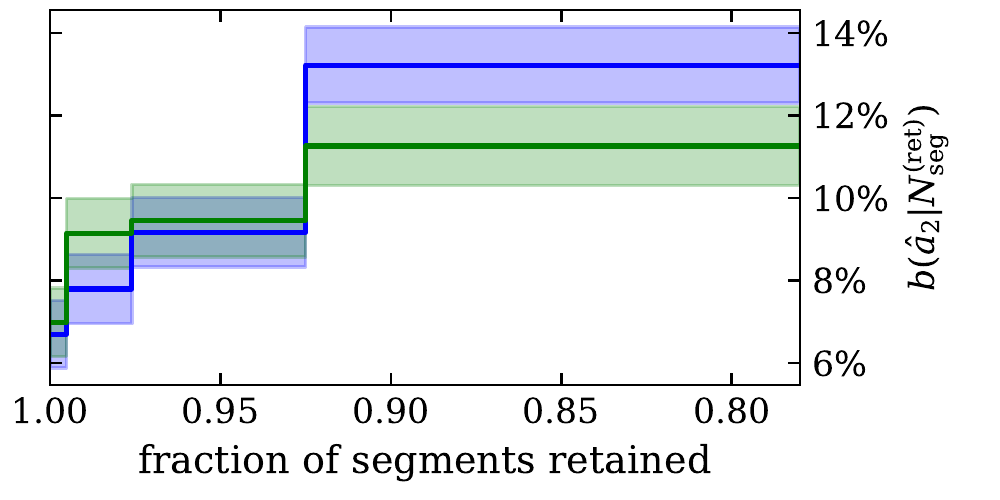}
        \includegraphics[width=1.0\textwidth]{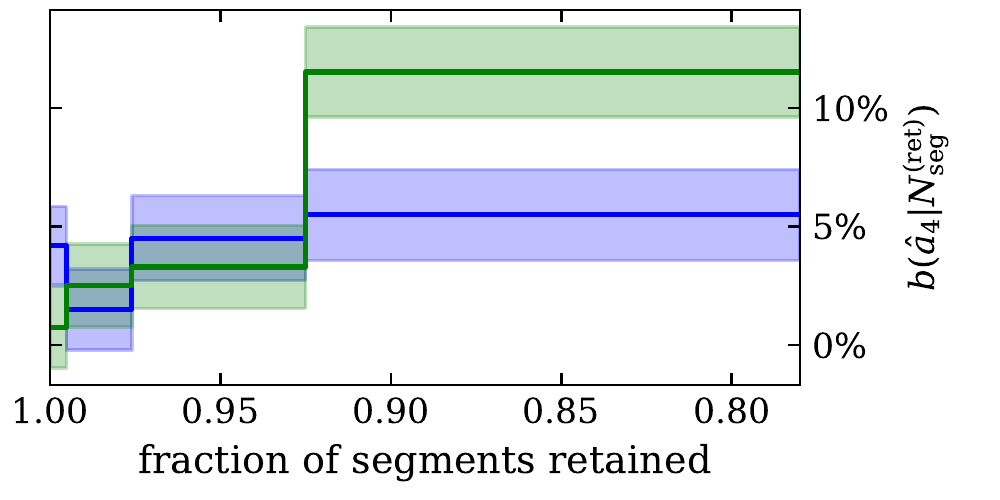}
    \end{minipage}
    \caption{
        (\emph{left}) Gaussian kernel density estimates for the distributions of our estimators for (\emph{top}) $\hat{a}_1$, (\emph{middle}) $\hat{a}_2$, and (\emph{bottom}) $\hat{a}_4$ conditioned on $N_\mathrm{seg}^{(\mathrm{ret})}=10^3$ with $a_0/\lambda_n = 2\%$ when I consider (\emph{orange}) all data ($N_\mathrm{seg}^{(\mathrm{ret})}=10^3$ axiomatically), (\emph{blue}) the retained subset ordered by counts, (\emph{green}) the retained subset of randomly ordered segments, and (\emph{black}) the true values.
        I shade the distributions up to their mean.
        The ordered and unordered subsets' distributions are nearly identical, both with means that are shifted to larger $\hat{a}_m$ than when I use all data.
        (\emph{right}) Corresponding estimates of the relative bias in $\hat{a}_m$ binned over $N_\mathrm{seg}^{(\mathrm{ret})}$: $b(\hat{a}_m|N_\mathrm{seg}^{(\mathrm{ret})})$ (Eq.~\ref{eq:bias}).
        Shaded regions approximate 1-$\sigma$ uncertainty in the bias due to the finite number of experiments.
        Bins are constructed to contain approximately equal numbers of experiments.
        The bias monotonically increases as the fraction of retained data decreases, and is always \result{$\gtrsim 7\%$} for $\hat{a}_1$ and $\hat{a}_2$.
    }
    \label{fig:persistent a_m bias}
\end{figure*}

In order to better understand the persistent bias in $\hat{a}_m$ introduced by H-test maximization, I consider the distribution of $\hat{a}_m$ conditioned on $N_\mathrm{seg}^{(\mathrm{ret})}$ in Fig.~\ref{fig:persistent a_m bias}.
I expect any bias to grow monotonically as I retain smaller fractions of the data based on the joint distributions in Fig.~\ref{fig:correlations}.
By estimating the distributions $p(\hat{a}_m|N_\mathrm{seg}^{(\mathrm{ret})}=10^3)$, I quantify the smallest the bias could be.
Fig.~\ref{fig:persistent a_m bias} shows this when $a_0/\lambda_n = 2\%$.
I see that both the ordered and unordered subsets demonstrate nearly identical distributions, and that these distributions are biased towards larger $\hat{a}_m$ than when I always use all the data.
That is, the mean of the distribution is larger than it would be if I always included all the data.
I define the relative bias as the ratio of expectation values
\footnote{My estimator for $\hat{a}_m$ is positive semi-definite and its distribution is skew right. It therefore predicts larger values on average than the true injection, even when I always include all the data. This effect is larger for weaker signals. For this reason, I define the size of the biases introduced by H-test maximization by normalizing them against the mean obtained when I always use all the data. This avoids overstating the effect, as might have been the case if they were normalized by the injected value.}
\begin{equation}\label{eq:bias}
    b(\hat{a}_m|N_\mathrm{seg}^{(\mathrm{ret})}) = \frac{\left<\right.\hat{a}_m|N_\mathrm{seg}^{(\mathrm{ret})}\left.\right>}{\left<\hat{a}_m\right>_\mathrm{All\,Data}} - 1
\end{equation}
and estimate the size of the bias as a function of $N_\mathrm{seg}^{(\mathrm{ret})}$ in Fig.~\ref{fig:persistent a_m bias}.

Put another way, even if the H-test maximization procedure instructs one to retain all the data, the resulting estimator is still biased relative to what one would obtain if they always included all the data.
As such, I conclude that H-test maximization is \emph{guaranteed} to introduce a bias in $\hat{a}_m$, regardless of whether any data are actually discarded.

I find that the mean of our estimators are \result{$\approx 7\%$} larger than when I always include all the data with $N_\mathrm{seg}^{(\mathrm{ret})}=10^3$ and $a_0/\lambda_n = 2\%$.
The bias grows as $N_\mathrm{seg}^{(\mathrm{ret})}$ decreases, and I find biases of \result{$\gtrsim 10\%$} when I retain 90\% of the data and \result{$\mathcal{O}(30\%)$} when I retain only half the data.

\subsection{Scaling of Bias with Signal Strength}
\label{sec:scaling of bias with signal strength}

The size of the bias scales strongly with the fraction of data that is retained ($N_\mathrm{seg}^{(\mathrm{ret})} / N_\mathrm{seg}$).
Typically, I find that maximizing $H$ for data containing larger injected signals tends to retain larger fractions of the data.
This makes sense, as larger signals are more easily detected in individual segments, and therefore it is more likely that each additional segment will increase $H$.
Conversely, maximizing $H$ for data with weak signals tends to discard larger fractions of data and introduce larger biases, on average.

\begin{figure}
    \includegraphics[width=1.0\columnwidth]{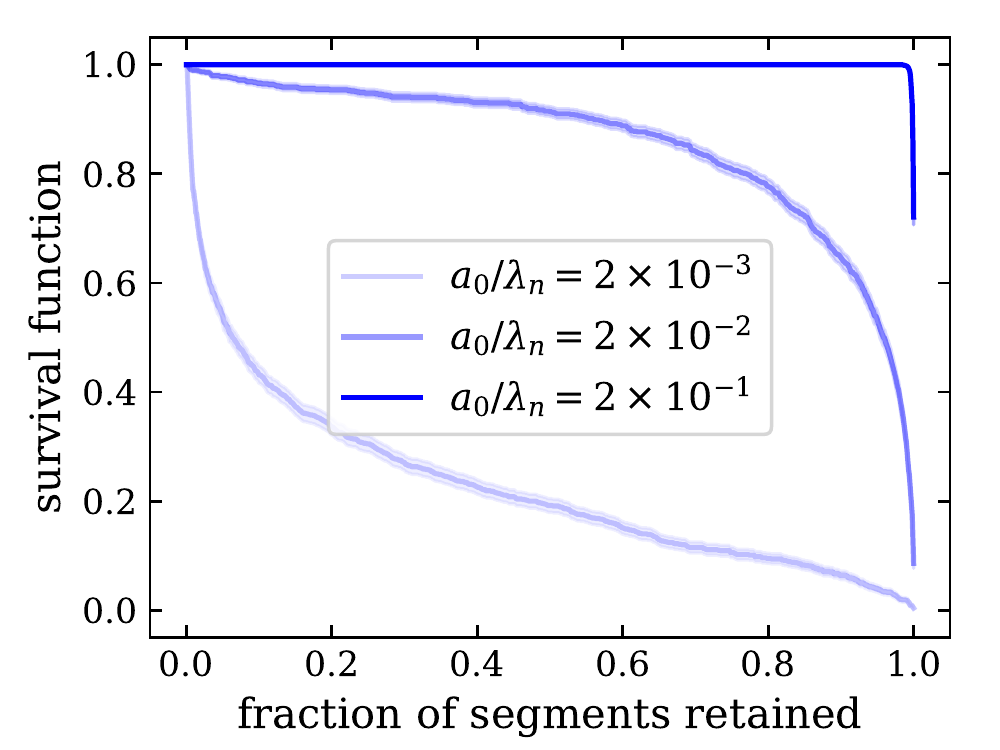}
    \caption{
        Survival function vs. the fraction of segments retained by the H-test maximization procedure.
        Shaded regions approximate 1-$\sigma$ uncertainty from the finite number of experiments.
        More segments are retained for louder signals, on average.
    }
    \label{fig:Nret distributions}
\end{figure}

Fig.~\ref{fig:Nret distributions} shows the distributions for the retained fraction of segments for different signal strengths.
Datasets with louder signals suffer less from the biases introduced by the data-selection process simply because one is less likely to discard large fractions of the data.
Nonetheless, biases are always present as the procedure still discards at least some of the data some of the time, regardless of the signal strength.

Furthermore, the type of persistent bias shown in Fig.~\ref{fig:persistent a_m bias} occurs for all signal strengths, but it is larger for weaker signals.
My simplified model predicts a bias of \result{$\approx 7\%$} when $N_\mathrm{seg}^{(\mathrm{seg})}=10^3$ and $a_0/\lambda_n = 2\%$.
This bias is exacerbated when $N_\mathrm{seg}^{(\mathrm{ret})}$ decreases.
The minimum bias is also larger when the signal is weaker, closer to \result{$\mathcal{O}(50\%)$} when $a_0/\lambda_n = 0.2\%$.

\subsection{Maximization Alters the Null Distribution}
\label{sec:maximization alters the null distribution}

\begin{figure*}
    \begin{minipage}{0.5\textwidth}
        \begin{center}
            \includegraphics[width=1.0\textwidth]{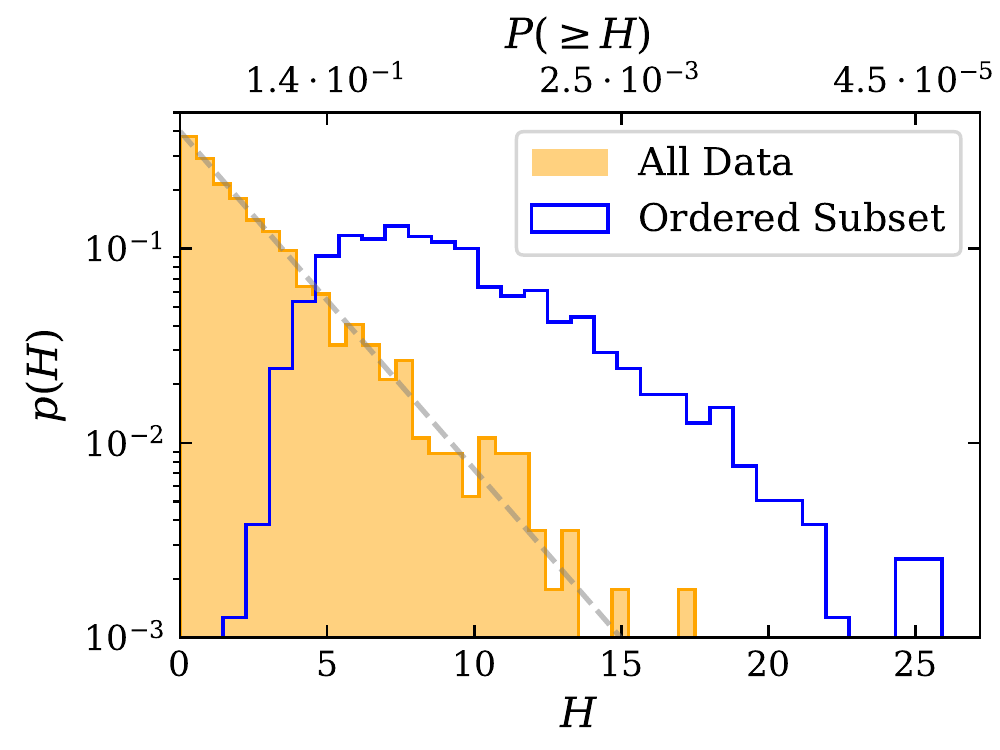} \\
            \vspace{-3.0cm}
            \huge{$\tau a_0 = 0$}
        \end{center}
    \end{minipage}
    \begin{minipage}{0.5\textwidth}
        \begin{center}
            \includegraphics[width=1.0\textwidth]{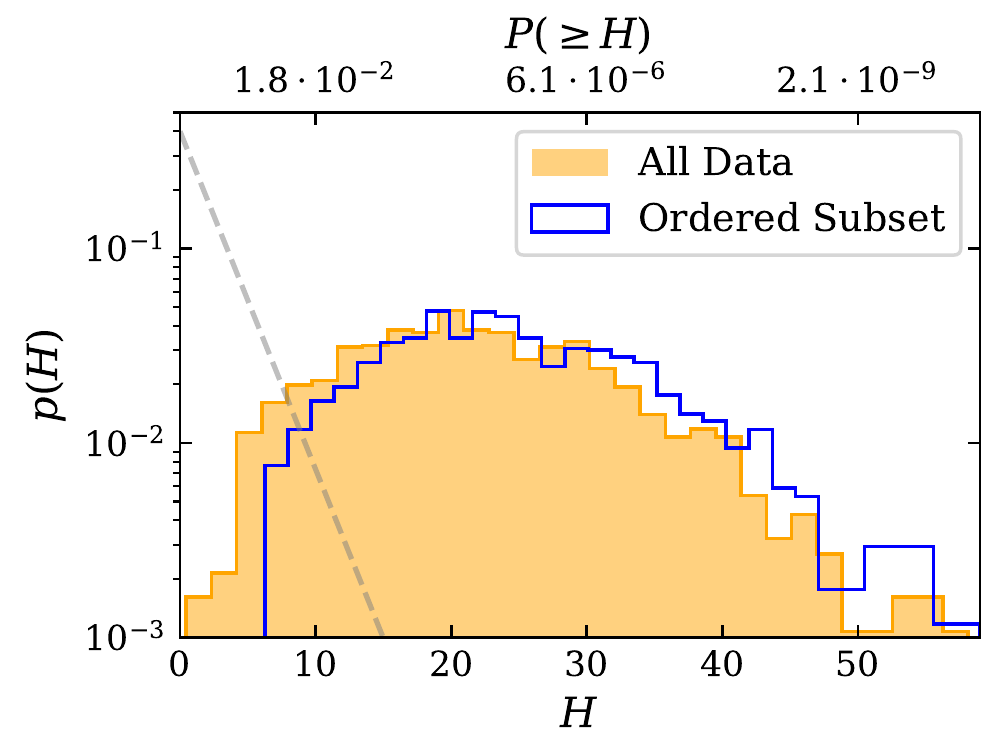} \\
            \vspace{-3.0cm}
            \huge{$\tau a_0 = 1$}
        \end{center}
    \end{minipage}
    \vspace{+2.0cm}
    \caption{
        Distributions of the H-test statistic ($H$) in (\emph{left}) the absence of a periodic signal ($a_0=0$) and (\emph{right}) with a moderate signal ($\tau a_0 = 1$, $a_0/\lambda_n = 2\%$) when we include (\emph{orange, shaded}) all available data and (\emph{blue, unshaded}) the subset retained by~\citet{Guillot:2019}'s procedure along with (\emph{grey dashed line}) the predicted distribution: $p(H)=0.4e^{-0.4H}$\citep{dejager:2010}.
        The top axis shows the corresponding nominal $p$-value under the assumption that all data are used (Eq.~\ref{eq:nominal pvalue}).
        The shift in $H$ can be of much more practical importance for weaker signals.
    }
    \label{fig:null distributions}
\end{figure*}

Finally, the maximization process changes the null distribution of the $H$ statistic.
\citet{dejager:2010} show that the $H$ statistic is exponentially distributed when all data are included and there is no signal present (Eq.~\ref{eq:nominal pvalue}).
Fig.~\ref{fig:null distributions} recovers this prediction within my idealized model.
However, the maximized H-test statistic produced by~\citet{Guillot:2019}'s data selection procedure follows a different distribution, even in the absence of a signal.
This means that significance estimates corresponding to values of $H$ obtained via~\cite{Guillot:2019}'s procedure cannot use the nominal distribution proposed by~\citet{dejager:2010}.
Indeed, in the absence of a signal, I find an average nominal $p$-value of \result{5.4\%} using the distribution of $H$ produced by \citet{Guillot:2019}'s procedure.
\result{10\%} of the time the nominal $p$-value is less than \result{0.3\%}.
Out of \result{$10^3$} experiments, the smallest nominal $p$-value observed by maximizing $H$ was \result{$3.2 \times 10^{-5}$}, nearly 2 orders of magnitude smaller than it should have been.

Therefore, one must take care when assessing the significance of possible signals using the maximized H-test.
While the value of $H$ is increased by the data selection procedure, this does not necessarily mean that it is more likely there is a signal present in the data.
Instead, the apparent increase in $H$ may be due only to the re-ordering and downselection of GTI's.

In general, there is still a shift toward larger $H$ when signals are present.
However, the relative importance is smaller for stronger signals.

\section{Discussion}
\label{sec:discussion}

While the results of Sec.~\ref{sec:toy model} suggest current data-selection procedures require reexamination, I reiterate that real NICER data are more complex than my idealized model.
The precision made available by NICER observations will continue to be invaluable for a broad community, and the goal of this Letter is only to help maximize the scientific value of those observations.

Nonetheless, the idealized model presented in Sec.~\ref{sec:toy model} shows that, even in the best-case scenario, the current data-selection procedure can lead to biases in the overall inferred count rate and the morphology of the inferred signal.
One might expect these effects to scale approximately as $N_\mathrm{count}^{-1/2}$ and therefore be less than 1\% for current NICER observations with $\gtrsim\mathcal{O}(10^4)$ counts.
However, I note that the total signal can be of the same order of magnitude for dim pulsars like J0740+6620 (signal amplitude is only a few percent).
The bias introduced may be a significant fraction of the true signal.

Briefly, I find that selecting a subset of data that maximizes the H-test detection statistic can bias estimators towards larger apparent modulations with phase, discard larger fractions of data for weaker signals, and change the H-test's null distribution, rendering standard significance estimates overly optimistic.
None of these effects are present if all available data are always used, and they are all directly attributable to the H-test maximization procedure.

I discuss these effects in the context of current NICER data in Sec.~\ref{sec:impact} before speculating on possible remedies in Sec.~\ref{sec:next steps}.
I again note that, while I focus only on the selection of GTI's,~\citet{Guillot:2019} also optimize the H-test over energy bands.
Further consideration should be given to the possible impact of optimization over both the energy band and GTI's.

\subsection{Impact on Published NICER Observations}
\label{sec:impact}

I again stress that real NICER data are more complicated than my idealized model.
While it would be foolhardy to attempt anything but the roughest error propagation from my model to the actual NICER results, it is still worth pointing out a few things.

\subsubsection{J0740+6620}
\label{sec:impact J0740}

First, it is worth noting that~\citet{wolff:2021} state, ``[t]he GTI selection is particularly effective for faint pulsars with count rates $\sim 0.01$ c/s such as PSR J0740+6620.''
This matches my observation that the bias and the increase in the H-test significance are more important for small signals.
\citet{wolff:2021} also states that the total set of data for J0740+6620 originally had 574,405 events between 0.31--1.22 keV and the downselected set had only 521,004 events between 0.31--1.18 keV.
This corresponds to a retained fraction of \result{$\approx 90.7\%$} of the counts.
Assuming the signal is only a few percent of the noise (like in Fig.~\ref{fig:correlations}), then this might corresponds to a bias on $\hat{\lambda}_0$ that is roughly 50\% of the size of the true signal's mean rate $a_0$.
However, some (or all) of this bias could be absorbed by the noise model (smaller $\hat{\lambda}_n$).

Indeed, within the context of J0740+6620, XMM-Newton~\citep{XMM-Newton} observations were used to estimate $a_0$ separately from the NICER data.
That is, assuming completely separate noise models for XMM and NICER, the analysis used historical XMM blank-sky observations to estimate the energy-dependent noise rate for XMM ($\lambda_n^{(\mathrm{XMM})}$) and observations of J0740+6620 to estimate the total count rate (signal and noise) from the pulsar ($\lambda_0^{(\mathrm{XMM})}$).
Assuming there are no biases in XMM data's selection, this should provide an unbiased estimate for the mean signal rate $a_0^{(\mathrm{XMM})} = \lambda_0^{(\mathrm{XMM})} - \lambda_n^{(\mathrm{XMM})}$.

While~\citet{Guillot:2019}'s data selection may still bias NICER's estimate of the total mean count rate $\lambda_0^{(\mathrm{NICER})}$, that bias could be completely accommodated by the flexible noise model assumed within the NICER analysis.
That is, the bias in the mean count rate observed in Fig.~\ref{fig:correlations} can be absorbed in $\hat{\lambda}_n^{(\mathrm{NICER})}$ without affecting the constraint on $a_0$ from XMM.
As such, one may suspect that the main impact of~\citet{Guillot:2019}'s procedure could be the bias towards larger $\hat{a}_{m}$ evident in Figs.~\ref{fig:correlations} and~\ref{fig:persistent a_m bias} rather than the bias towards lower $\hat{\lambda}_0$.

Assuming my simulation with $a_0/\lambda_n = 2\%$ is characteristic of the observed signal strength for J0740+6620, I estimate possible biases in $\hat{a}_m$ of \result{$\mathcal{O}(10\%)$} when 90\% of segments are retained.
If XMM data provides an unbiased estimate of $a_0$, one then can expect an $\mathcal{O}(10\%)$ bias in the relative modulation depth: $a_m/a_0$.
In the context of our simplified model, the bias for $\hat{a}_2$ is larger than $\hat{a}_1$, suggesting that the primary signal harmonic may be more biased than other harmonics.

However, we also note that the mapping from $a_m/a_0$ to the compactness is not linear, and small increases in the compactness (smaller radii) can quickly flatten the light curve (much smaller $a_m/a_0$), whereas low compactness (large radii) may not be as easily constrained (only slightly larger $a_m/a_0$).
As such, my estimate of a \result{$\mathcal{O}(10\%)$} effect on the relative modulation depth may not correspond to a shift in the lower bound on J0740+6620's radius of similar size.
Nonetheless, it would be of interest to determine exactly how large this shift could be, as the difference between a $2\,M_\odot$ star with a radius of, e.g., 12.5 km vs. 13.0 km ($\approx 4\%$ shift) could have nontrivial implications for the dense matter EoS, e.g., the existence of strong phase transitions with large onset densities or nearly causal sound speeds at high densities~\citep{legred:2021, Drischler:2021, Drischler:2021bup}.

An early version of this Letter was shared with the NICER lightcurve working group, and they reanalyzed J0740+6620 using all available GTI's and an essentially unchanged energy band (\result{0.30-1.23} keV vs. \result{0.31-1.22} keV), finding an \result{$\mathcal{O}(1\%)$} increase in the lower bound on the radius compared to the published results~\citep{ColeAndAlex}.
This is in line with the predictions of my idealized model, which suggests changes in $\hat{a}_m$ anywhere between \result{$-5\%$ and $+20\%$} are possible when one compares the ordered subset of GTI's to the result using all the data for a single experiment.
Nonetheless, just because the constraints obtained from a single observation are not strongly affected by \citet{Guillot:2019}'s procedure, this does not mean that no bias exits.
Observations of other pulsars, particularly dim stars, may still be significantly affected.

\subsubsection{J0030+0451}
\label{sec:impact J0030}

Observations of J0030+0451 may be less affected by such issues.
At first glance, J0030+0451 has much larger modulations~\citep[$\gtrsim 30\%$ peak-to-peak, see][]{Bogdanov:2019} than J0740+6620~\citep[$\lesssim 8\%$ peak-to-peak, see][]{wolff:2021}.
Therefore, one expects the impact of maximizing the H-test significance to be less for J0030+0451.
What's more, while selection cuts were placed on J0030+0451 data based on the count rate in 16 sec segments, the threshold was not chosen to maximize the H-test significance~\citep[see discussion in][]{Bogdanov:2019, miller:2019}.
Indeed,~\citet{miller:2019} discards any segments with count rates above \result{3 Hz}, noting that the average count rate is \result{0.7 Hz}.
For 16 sec segments, this threshold corresponds to approximately 11-$\sigma$ fluctuations in the number of counts per segment.
Such a selection may still introduce a bias, but it is likely vanishingly small and unimportant for all practical considerations.

\subsection{Possible Paths Forward}
\label{sec:next steps}

Fundamentally, these biases occur because the likelihood does not reflect the true data generation process.
Selecting a subset of data (that is not chosen completely randomly) will inevitably bias the inference as the unaltered Poisson likelihood does not account for the additional data-selection process.

One possible solution is to simply use all data that were recorded as was done for J0740+6620 in~\citet{ColeAndAlex}.
However, as I have said several times, real NICER data are more complicated than my idealized model, and the actual background can vary over time.
Using all the data may simply exchange one source of bias (selecting a subset of data) for another (misspecification of the noise model) or increased statistical uncertainty from segments with very large count rates.
As such, one may still wish to implement some sort of data-selection procedure based on the counts in short segments.

More work will be needed to determine the extent of the selection effects described herein for more complicated data (e.g., variable noise rates) and in the presence of optimization over both GTI's and energy channels.
A thorough understanding of such effects may allow them to be modeled and accounted for within future analyses.

\begin{acknowledgements}

\begin{center}
  -- \emph{Acknowledgments} --
\end{center}

I am very thankful for the many detailed discussions with Cole Miller during the preparation of this manuscript.
I also thanks the broader NICER Collaboration for entertaining the implications of this idealized model for the much more complicated analyses they perform that enable exceptional science with X-ray observations.

Research at Perimeter Institute is supported in part by the Government of Canada through the Department of Innovation, Science and Economic Development Canada and by the Province of Ontario through the Ministry of Colleges and Universities.
I also thank the Canadian Institute for Advanced Research (CIFAR) for support.
This work was performed in part at the Aspen Center for Physics, which is supported by National Science Foundation grant PHY-1607611.

This work was made possible by the \texttt{numpy}~\citep{numpy}, \texttt{scipy}~\citep{scipy}, and \texttt{matplotlib}~\citep{matplotlib} software packages.
All code used to generate synthetic data within this study is publicly available at $\left<\right.$\url{https://github.com/reedessick/htest-maximization}$\left>\right.$.

\end{acknowledgements}


\bibliography{refs}

\end{document}